# Development of bed forms due to waves blocked by a counter current

**Debasmita Chatterjee[1], Subir Ghosh[2], B. S. Mazumder[3] and K. Sarkar[4]**

Fluvial Mechanics Laboratory, Physics and Applied Mathematics Unit
Indian Statistical Institute, 203 B. T. Road, Kolkata 700108, India

[3]Present address: Department of Aerospace Engineering and Applied Mechanics
Indian Institute of Engineering Science and Technology (IIEST), Shibpur 711103, India
Department of Civil Engineering, I. I. T., Bombay, Mumbai 400 076, India

ABSTRACT: Experiments are conducted in a laboratory flume on the propagation of a surface wave against unidirectional flow with a sediment bed. This paper presents the spatial variation of bed forms induced by the wave-blocking phenomenon by a suitably tuned uniform fluid flow and a counter-propagating wave. The occurrence of wave-blocking is confirmed by finding a critical wave frequency in a particular flow discharge in which the waves are effectively blocked and is established using the linear dispersion relation. The novelty of this work is to identify the wave-blocking and its influence on the development of bed forms over the sediment bed. Interesting bed form signatures are observed at a transition of bed forms in three zones, with asymmetric ripples having a steeper slope downstream face induced by the incoming current, followed by flat sand bars beneath the wave-blocking zone and more symmetric ripples below the wave-dominated region at the downstream. This phenomenon suggests that the sediment bed is segmented into three different regions of bed geometry along the flow. The deviations of mean flows, Reynolds stresses, turbulent kinetic energy, and power spectral density due to the wave-blocking phenomenon are presented along the non-uniform flow over sediment bed. The bottom shear stress, bed roughness and stochastic nature of the bed form features are also discussed. The results are of relevance to engineers and geoscientists concerned with contemporary process as well as those interested in the interpretation of palaeoenvironmental conditions from fossil bed forms.
.

**[1] Email: debasmitachatterjee424@gmail.com, [2] Email: subirghosh20@gmail.com, [3] Email: bsmazumder@gmail.com (Corresponding author), [4] Email: soundofphysiks@gmail.com**

# Introduction

The combined wave-current flows govern many physical processes of interest to oceanographers and engineers working in coastal environments. For example, sand ripples are formed due to the synergetic effect of waves and currents on the sediment bed. In coastal and estuarine regions, waves and currents coexist to create a very effective agent for sediment transport. Several theoretical and experimental investigations had been carried out over the last few decades into counter- or co-propagating waves with turbulent flows over smooth/rough beds (Grant and Madsen, 1979; Brevik and Aas, 1980; Kemp and Simons, 1982, 1983; Mathisen and Madsen, 1996a, b; Umeyama, 2005, 2009; Silva et al. 2006; Mazumder and Ojha 2007; Ojha and Mazumder, 2010; Olabarrieta et al., 2010; Singh et al., 2015).

Brevik and Aas (1980) reported results of both following and opposing shear flows with surface waves over ripple beds to study the mean velocity and friction factors. Kemp and Simons (1982) described an experimental study for combined wave-current flows over smooth and rough boundaries, where a wave co-propagated with a flow; and a considerable increase in bed shear stress was observed at the rough boundary. Subsequently, for counter propagating waves, Kemp and Simons (1983) observed that the near-bed velocities for the rough-bed were reduced, while near-bed turbulence intensities for both rough and smooth surfaces were increased by the presence of surface waves. A simple model for near-bottom combined wave-current flows over a moveable sediment bed was developed by Glenn and Grant (1987), using unsteady conservation of momentum and sediment mass equations, which were coupled through an eddy diffusivity closure scheme. Yu and Mei (2000) described a quantitative theory for the formation of sand bars under surface waves, using the process of forced diffusion. The development of bed forms induced by solitary waves in shallow water over an initially flat sand bed was studied by Marin et al. (2005) in a laboratory flume. Umeyama (2005, 2009) performed a series of experiments in an open channel flow with only a current, waves following a current and waves against a current over a rigid flat bed, and analysed the mean velocity and phase-averaged Reynolds stresses. Moreover, they reported that the wave attenuation increases, which results in the loss of wave energy for waves against current, and hence, an increase in the mean flow near the water surface. On the other hand, there is less wave attenuation, less loss of energy and a reduction in the mean flow near the water surface for the waves following the current.

Despite all these studies, no such experimental studies are directly performed to examine the development of bed forms associated with turbulence due to wave-blocking on a sand bed when a wave counter-propagates with a current. The effect of wave against flow induces the wave-blocking when the incident wave is diminished by a sufficiently strong current. If the current is strong enough, the group velocity goes to zero, which leads to the wave-blocked phenomenon (Chawla and Kirby, 2002). Hence, a region is created beyond which the waves are absent and only the background flow of fluid remains. At a river mouth, for example, the river flow blocks the incoming sea waves. This is a common phenomenon at the entrance of tidal inlets, where the tidal currents are strong. Chawla and Kirby (2002) performed a series of experiments to comprehend the dynamics of the interaction between waves and strong opposing currents for wave-blocking, using the simplified wave action conservation equation proposed by Bretherton and Garrett (1968) and the linear dispersion relation. They studied the energy dissipation due to monochromatic and random wave breaking at or before blocking points. Recently, Chatterjee et al. (2018) experimentally studied the turbulence features of the wave-blocking phenomenon, where the opposing current gradually blocked the primary wave energy, i.e., where the wave-propagation velocity vanished. They showed that there was an evolutionary change in the flow with three segmented regions, i.e., the flow at the upstream, blocking at the mid-stream and waves in the downstream. Some studies are available on regular and irregular waves over muddy beds with following or opposing currents (An and Shibayama, 1994; Zhao et al., 2006; Soltanpour et al. 2007, 2014). An and Shibayama (1994) experimentally investigated the regular wave-current interaction on the fluid mud layer and found higher wave dissipation rates and mud transport velocities due to an opposing current. Soltanpour et al. (2014) performed a series of experiments to study the attenuation of waves with current over the mud layers, using the simplified wave action conservation equation; they also studied the wave spectra for a wave only, opposing and following currents. The effect of combined wave-current flows is relevant to understand the sediment bed form features, when wave-blocking occurs over a sand bed.

The present paper focuses on the development of bed forms associated with the turbulence due to the wave-blocking phenomenon for counter-propagating waves with a current. An attempt has been made to identify the wave-blocking condition from an interaction of waves with a counter current on the sediment bed, using the dispersion relation of monochromatic waves, and the effect of wave-blocking with the associated turbulence on the sediment bed is examined. This study provides interesting signatures in the bed forms with asymmetrical ripples having steeper slope downstream faces induced by the incoming flow, followed by a flat sand bed beneath the wave-blocking region and more symmetrical ripples below the wave-dominated region downstream. The key turbulence parameters are evaluated at different regions of non-uniform flow along the flume, which characterize the formation of three distinct bed form geometries. The velocity power spectra and exceedance probabilities of the bed elevations are estimated for all three segmented regions. This study may be useful for the engineers that are concerned with the dynamics of flow and bed forms (An and Shibayama, 1994; de Swart and Zimmermam, 2009) and for geoscientists interested in palaeoenvironmental conditions (Kolmogorov, 1951; Bhattacharya and Giosan, 2003; Sengupta, 2007).

## Experimental setup and data collection

### Test channel

Experiments were carried out in a specially designed re-circulating flume (Sarkar et al., 2016) at the Fluvial Mechanics Laboratory, Physics and Applied Mathematics Unit, Indian Statistical Institute, Kolkata. Both the experimental and re-circulating channels of the flume are of the same dimensions (10 m long, 0.5 m wide and 0.5 m deep). Perspex windows with a length of 8 m provide a clear view of the flow (Figure 1). The water depth is kept constant at $H = 25$ cm for all experiments and the hydraulic slope of the flume is of the order of 0.001. An electromagnetic flow meter on the outlet pipe facilitates continuous monitoring of discharge. The cross-sections of both inlet and outlet pipes were smaller than the wetted cross-section of the flume (Figure 1a of Chatterjee et al. 2018).

### Wave-maker

A plunger-type wave-maker is mounted at the downstream end of the flume at a distance 8.5 m from the source to generate surface waves against the current. A triangular shaped six-inch cylinder closed at both ends, similar to the wave-maker of Euve et al. (2015), is partially submerged in the water (Figure 1). When the spindle is rotated using a motor of variable speed, the cylinder can move up and down with variable frequency to generate waves against the current. The amplitude of oscillations is fixed at 11 cm. The wave-maker is fixed with a Variac (variable resistor) to control the frequency of oscillation, as desired. The descriptions of the plunger type wave-maker with a control box are available in Mazumder and Ojha (2007); Chatterjee et al. (2018). The calibration is performed using a tachometer for frequency variation. A wave-absorber is placed further downstream behind the wave-maker.

## Sediment bed

The flume was laid with a smooth flat sand bed, with at thickness $h'$ (= 10 cm), length of 8 m and a width of 0.5 m. The median particle diameter $d_{50}$ was 0.25 mm with a geometric standard deviation $\sigma_g$ = 0.685. The specific gravity of the sediment was 2.65. Here, the mean flow depth was $h = H - h' = 15$ cm, which was above the flat sand bed (Figure 1). The coordinate system of measurement was as follows: x positive downstream along the flow, y-axis transverse to the flow, and z positive upward. The origin (0, 0, 0) was at the inlet source near the honeycomb.

**Figure 1**

**Figure 2(a, b, c)**

## Data collection procedures

The study was performed systematically in three steps, as follows:

*Step-1*: To check the fully developed flow over the flat rigid bed, when the flow achieved an equilibrium state after approximately 60 minutes of operation (Chatterjee et al. 2018), the instantaneous velocity data were collected from three different locations. A 3-D Micro-acoustic Doppler velocimeter (ADV) was used to collect the velocity for 180 sec at a sampling rate of 40 Hz along the centreline of the flume. Two different flow discharges ($Q$ = 0.0197 and $0.0322\,m^3/\sec$) with corresponding Reynolds numbers, i.e., $Re$ = 0.62x10$^5$ and 1.02x10$^5$, were used to confirm the fully developed flow at the experimental section (Table 1a), which was 4 m from the entry zone. The data were collected from the lowest level, i.e., 0.3 cm above the flat rigid bed to the highest level, i.e., approximately 18 cm; no velocity data were measured near the free surface. An electromagnetic discharge meter was fitted to the inlet and outlet pipes to facilitate continuous monitoring of the flow. In the present case, the flow discharge was computed at the sampling station using the depth-averaged velocity and wetted cross-section. The computed flow discharge ($Q_{cal}$) was approximately 1.60 times greater than of the recorded discharge ($Q$) using an electromagnetic discharge meter (Step-1, Table-1). Detailed descriptions of the experimental flume with inlet and outlet pipes are available in Chatterjee et al. (2018). The recorded discharge ($Q$) is referred throughout the paper for presentation.

*Step-2:* There was a unidirectional flow of discharge $Q$ (Table 1b) over the flat sand bed (Figure 2a) of median size $d_{50}$ = 0.25 mm. The fluid velocity was chosen in such a way that the bottom shear stress was higher than the threshold value for sediment movement. After a certain time of operation, the bed form developed regular asymmetrical ripples with steeper slopes at the downstream faces (Figure 2b) over the sand bed surface. The bottom shear stress was characterized by the dimensionless Shields parameter $\tau_0^*$ and can be approximated as follows:

$$\tau_0^* = \frac{\tau_0}{(\rho_s - \rho)\, \mathrm{g}\, \mathrm{d}_{50}}, \tag{1}$$

where $\tau_0 = \rho g h_R s$ with $h_R = \dfrac{bh}{b + 2h}$ is the hydraulic radius, $b$ is the width of the flume, $h$ is the averaged water depth, $s$ the hydraulic slope of the flume (of order 0.001), $\rho_s$ is the sediment density, $\rho$ is the fluid density and g is the acceleration due to gravity. In the present case, the Shield parameter $\left(\tau_0^*\right)$ is approximately 0.23, which is greater than that of the incipient motion. However, the computation is also made using $\tau_0 = \rho u_*^2$ in Eq. (1), where the friction velocity $u_*$ is approximately 3.2 cm/s at flow region over sand bed during the wave-blocking condition, which confirms the value greater of incipient motion. When the bed forms achieved the near equilibrium state, the fluid flow was stopped, and the bed elevations were recorded in still water using the Micro-ADV at every 2 cm distance from upstream to downstream covering a distance of approximately 3.5 m along the centreline of the flume. It may be mentioned here that ADV can also be used to record the bed elevation from a particular level of water depth. Similar experiments were performed over the sediment bed of grain size $\mathrm{d}_{50} = 0.35$ mm using identical or higher discharges.

**Table 1**

*Step-3:* To identify the wave-blocking condition from waves propagating against a current on the given asymmetrical sediment rippled bed, several test runs were worked out superposing the opposing waves of different frequencies ($\omega$) on the given discharge ($Q$). Thus, for a discharge $Q$, the counter propagating frequency of the wave generated from the wave-maker was regulated in such a way that the wave-blocking occurred at a location on the free surface. It was observed that there was a wave-blocking region for a given discharge ($Q$) and a critical frequency ($\omega_c$). Similarly, the experiments were repeated for a range of discharges and wave frequencies to identify the effect of wave-blocking over the sediment bed (Table-1c). It was observed that the increase of the flow discharge $Q$ led to a decrease in the critical frequency ($\omega_c$) for wave-blocking. For a lower frequency of wave opposing the flow, the amplitude was relatively high for the given discharge $Q$ and it decreased around the blocking region when frequency was increased. The amplitude approached zero at a point when the frequency was sufficiently high and, hence, the blocking condition was reached. In the present case, when the waves opposed the flow, the incidence wavelengths were seen to be reduced as when they approached the blocking region and diminished to millimetres or less in magnitude due to the loss of wave energy.

The predetermined critical frequency $\omega_c$ of the opposing waves over the sediment bed was superimposed with the given unidirectional flow to examine the amendment of the existing asymmetric bed forms developed in step-2. In the experiment, the incoming waves of critical frequency ($\omega_c$) were blocked over the sediment bed when the counter current became sufficiently strong (Figure 2 (c), Table-1c). The amplitudes of a given frequency of the counter propagating waves travelling along the water surface were recorded by a digital camera, particularly at the wave-blocking and wave-dominated regions. After a certain time (240 minutes), when the transformed bed forms achieved nearly in equilibrium state, the instantaneous velocity data were recorded using ADV for 180 sec at a sampling rate of 40 Hz at several vertical levels starting from the lowest level of 0.5 cm above the sand bed to the highest level at approximately 9.5 cm for the Reynolds number *Re* and the Froude number *Fr*. Below the level of 0.5 cm, the measurements of the velocity data were undetected, because the particle–particle interactions were very high and the sediment bed surface was moving as bed load sheet. The velocity data were collected vertically from five different locations along the flume centreline, namely, in the flow region, the wave-blocking region and the wave-dominated region, to quantify the deviations of the mean velocities, friction velocities, spectral densities and associated turbulence. Later, the flow was stopped and a process similar to step-2 was utilized to record the variations of bed elevation at every 2 cm distance from upstream to downstream along the centreline of the flume, and photographs of the modified bed forms were taken along the flume for analysis. Similarly, the experiments were repeated for a range of discharges and wave frequencies to study the effect of wave-blocking on the corresponding bed form development.

## Data analysis and results

## Mean flows over a rigid flat bed

The collected instantaneous velocity data were processed to remove spikes using the phase space threshold despiking method described by Goring and Nikora (2002). The instantaneous velocities ($u$, $v$, $w$) in the Cartesian coordinate system ($x$, $y$, $z$) include a mean and a fluctuating component given by the following:

$$u = \bar{u} + u'; \quad v = \bar{v} + v'; \quad w = \bar{w} + w' \tag{2}$$

where $\bar{u}$, $\bar{v}$, $\bar{w}$ = time-averaged velocities along ($x$, $y$, $z$) directions; $u'$, $v'$, $w'$ = the fluctuations of velocity components $u$, $v$, $w$. The time-averaged stream-wise, lateral and bottom-normal velocity components ($\bar{u}$, $\bar{v}$, $\bar{w}$) are defined as follows:

$$\bar{u} = \frac{1}{n}\sum_{i=1}^{n} u_i , \quad \bar{v} = \frac{1}{n}\sum_{i=1}^{n} v_i , \quad \bar{w} = \frac{1}{n}\sum_{i=1}^{n} w_i \tag{3}$$

where n is the number of observations ($n = 7200$) during the sample period of 180 sec. It was observed that there was no change among the three locations, confirming the fully developed flow along the test section. The dimensionless stream-wise mean velocity profiles over the flat rigid bed were found to follow a log-law with a coefficient of regression $R^2 \approx 0.95$ at all the measuring stations as follows:

$$\frac{\bar{u}}{u_*} = \frac{1}{\kappa}\ln\frac{z}{z_0}, \quad z_0 = \frac{k_s}{30} \tag{4}$$

where $\kappa$ is the von Karman constant (0.4), $k_s$ is the equivalent bed roughness (equals 0.025 cm), and $u_*$ is the friction velocity determined from the log-law. The vertical velocity was found to be almost zero throughout the depth of the flow. The universal functions of normalized turbulence intensities in the stream-wise and vertical directions above the rigid flat bed are computed against the normalized vertical distance $z/H$ (the profiles are not shown here). The results are found to be in good agreement qualitatively with the existing results. Therefore, a reasonable uniformity in the flow is observed under the given flow conditions. The respective values of friction velocities $u_*$ are 0.91 and 1.59 cm/sec, determined from the log-law, and the Froude numbers are 0.158 and 0.261.

## Wave-current interaction equations for wave-blocking

In the present analysis, the linear dispersion relation is used to study the wave-current interaction for the low amplitude waves. To study the wave-blocking phenomenon under an opposing current, it is convenient to use the wave action equation (Bretherton and Garrett, 1968), which describes the ratio of the current velocity and group velocity. The dispersion relation between frequency $\omega$ and wave vector $k$ for an incompressible, small amplitude gravity wave on a fluid surface is given by the following:

$$\omega^2 = gk \tanh(kH) \tag{5}$$

where $g$ and $H$ are the acceleration due to gravity and the depth of the fluid, respectively. We ignore the surface tension effects, as this is not significant in this case. To apply this dispersion relation in an interacting fluid current - gravity wave system, several points need to be taken care of. The first one is essentially the Doppler Effect, since waves on a moving medium are involved.

For a quick derivation, consider two coordinate systems $\vec{r} = (x, y, z)$, $\vec{r}' = (x', y', z')$, where the former (unprimed frame) moves with the uniform horizontal velocity $\vec{U}$ (of the fluid current) and the latter (primed frame) is the fixed laboratory frame. The two systems are related by Galilean transformations $\vec{r}' = \vec{r} + \vec{U}t$. A simple harmonic progressive wave train,

$$\psi = A \sin(\vec{k}.\vec{r} - \omega t) \tag{6}$$

in the moving frame will be transformed to the following:

$$\psi = A \sin(\vec{k}.\vec{r}' - \omega t) = A \sin(\vec{k}.(\vec{r} + \vec{U}t) - \omega t)$$
$$= A \sin(\vec{k}.\vec{r} - (\omega - \vec{U}.\vec{k})t) = A \sin(\vec{k}.\vec{r} - \sigma t) \tag{7}$$

where $\sigma = \omega - \vec{k}.\vec{U}$ is the intrinsic or moving frequency that is the frequency seen in a frame moving with the fluid velocity $\vec{U}$. This leads to a modification in the dispersion relation, following the researchers (Chawla and Kirby, 2002; Haller and Ozkan-Haller, 2002; Soltanpour et al., 2014), as follows:

$$\sigma^2 = (\omega - \vec{k}.\vec{U})^2 = gk \tanh(kH) \tag{8}$$

Differentiating $\sigma$ in Eq. (8) with respect to $\vec{k}$ gives the following:

$$\vec{c}_{ga} - \vec{U} = \vec{c}_g \tag{9}$$

where $\vec{c}_{ga} = \partial \omega / \partial \vec{k}$ is the group velocity vector in a stationary frame and $\vec{c}_g = \partial \sigma / \partial \vec{k}$ is the group velocity vector in a moving frame. If the wave propagates along the x direction, we have the following:

$$c_{ga} = U + c_g \qquad (9a)$$

The wave-blocking occurs at a point where the absolute group velocity vanishes, that is, at the blocking point $c_{ga} = 0$, Eq. (9a) gives the following:

$$U + c_g = 0 \qquad or \quad c_g = -U \qquad (10)$$

The conservation principle of wave action expressed by Bretherton and Garret (1968), is of the following form:

$$\frac{\partial}{\partial t}(\frac{E}{\sigma}) + \nabla.[\frac{E}{\sigma}(\vec{c}_g + \vec{U})] = 0 \qquad (11)$$

where E is the energy density and $c_g$ is the group velocity. In fact, for the steady state, Eq. (11), termed as the wave action equation, is the same as the radiation stress equation proposed by Longuet-Higgins and Stewart (1961). Therefore, dropping the time derivative term from (11) leads to the following:

$$\frac{E}{\sigma}(c_g + U) = Const. \qquad (12)$$

Furthermore, using the relation $E = \rho g H'^2 / 8$, where $\rho$ is the fluid density and $H'$ the wave height, we have the following:

$$\frac{H'^2}{\sigma}(c_g + U) = Const. \qquad (13)$$

To obtain the constant from (12), one has to fix a reference value for the no flow condition (i.e. $U = 0$). Thus, following Huang et al. (1972), the Eq. (12) is rewritten as follows:

$$\frac{E_0 c_{g0}}{\omega} = \frac{E(c_g + U)}{\sigma} \qquad (14)$$

where subscript $0$ refers to the no fluid flow condition with $\sigma$ is the angular wave frequency in the moving frame of reference and $\omega$ is the angular frequency.

Finally, differentiating Eq. (8) with respect to k, one can find the group velocity $c_g$ as follows:

$$c_g = \frac{1}{2}(1 + \frac{2kh}{\sinh 2kh})\frac{\sigma}{k} \qquad (15)$$

## Verification of wave-blocking over the sediment bed

The opposing waves of fixed frequency ($\omega_c$) identified from the wave-blocking condition were superimposed on the flow over the existing asymmetric rippled sand bed. The newly generated sand bed was compared with that of earlier bed forms developed in step-2. This process was repeated for different discharges ($Q$), and in each case, the similar transitional bed forms were noticed over the sand bed throughout the flume when the wave-blocking condition was reached. Following Dean and Dalrymple (2000), the wave characteristics were computed for different discharges ($Q$ = 0.0169, 0.0183, 0.0197, 0.0208 $m^3/\text{sec}$). Here, the mean flow depth is $h = H - h' = 15$ cm, which is above the sand bed $h' = 10$ cm thick. For critical frequencies $\omega_c$ = 1.23 Hz, 1.15 Hz and 1.10 Hz, the $kh > \pi$ indicates a deep-water wave, whereas for $\omega_c$ = 1.01 Hz, $kh < \pi$ indicates intermediate deep-water waves (Table 1c), where $k(= 2\pi/\lambda)$ is the wave number. When waves propagate against a current, the wave attenuation increases, resulting in loss of wave energy. On the other hand, when the waves propagate along a current, there is less wave attenuation, less loss of wave energy, and a corresponding decrease in velocity near the surface (Kemp and Simons, 1983). For discharge $Q = 0.0197 \, m^3/s$, we obtain the values of $c_{ga}$ = 30.8 cm/s, $U_{avg}$ = 14.197 cm/s, $c_g$ = 16.60 cm/s, and the ratio of depth averaged velocity $U_{avg}$ over the sediment bed and the group velocity $c_g$ is 0.855; and for the discharge Q = 0.0183 $m^3/s$, the value of $c_{ga}$ = 28.75 cm/s, $U_{avg}$ = 13.695 cm/s, $c_g$ = 15.055 cm/s, and the ratio of $U_{avg}$ and $c_g$ is 0.91. Therefore, it is revealed that the ratio is approximately one, indicating the formation of wave-blocking over the sediment bed. Hence, the condition of wave-blocking is confirmed from the present data.

## Mean velocities over the deformed sand bed

To study the velocity data collected vertically from five different horizontal locations, namely, at the flow region ($x/h$ = 28 and 42), the blocking region ($x/h$ = 44) and the wave region ($x/h$ = 47 and 50), the stream-wise and bottom-normal mean velocity in the respective locations are analysed, where $x$ is the longitudinal distance starting from $x = 0$ at the inlet source and $h$ is the mean flow depth above the sediment bed. Similarly, in the turbulent flow, the instantaneous velocity components ($u, v, w$) are defined as the sum of the respective mean velocity and the fluctuating velocity components, i.e., Eqs. (2) and (3).

When waves are combined with a current, the periodic disturbance produces sinusoidal motion in the velocity signals except near the boundary ($\frac{z}{h} < 0.2$). Therefore, it is important to introduce the phase-averaged quantities because the instantaneous velocity is modified by the mean velocity, the time-dependent velocity and the fluctuating velocity due to waves.

Accordingly, the instantaneous velocity components *u* and *w* usually take the following form (Umeyama, 2009):

$$\mathrm{u} = \overline{u} + \tilde{u}(t) + u' \tag{16}$$

$$\mathrm{w} = \overline{w} + \tilde{w}(t) + w' \tag{17}$$

where $\tilde{u}(t) = \langle u(t) \rangle - \overline{u}$ and $\tilde{w}(t) = \langle w(t) \rangle - \overline{w}$, with the over-bar denoting the mean velocity, the tilde (~) denoting wave-induced velocity, the prime denoting the fluctuating velocity, and the <> symbol denoting the phase-averaged velocity. The wave-induced velocity $\tilde{u}(t)$, which possesses a periodic nature, is obtained by subtracting the mean velocity $\overline{u}$ from the phase-averaged velocity $\langle u(t) \rangle$. The phase-averaged velocity $\langle u(t) \rangle$ is determined by computing the average over an ensemble of samples taken at a fixed phase in oscillation. The phase-averaged velocity is defined as follows (Umeyama, 2005; Singh et al. 2015):

$$\langle u \rangle = \left[ \frac{1}{\mathrm{N}} \sum_{n'=0}^{\mathrm{N}-1} u(t + n'T) \right] \tag{18}$$

where $n'$ = oscillation cycle number, $T$ is the wave period, N is the total number of oscillation cycles. The phase-averaged velocity $\langle u(t) \rangle$ can be extracted from the instantaneous velocity *u* by cross-correlation with a sinusoidal wave in phase with oscillation. Here, N = 50 sinusoidal wave cycles are reproduced for velocity time series at a measurement point. However, these wave cycles are not observed in the wave-blocking region or further upstream.

The stream-wise and bottom-normal mean velocity profiles in the respective locations are plotted against the normalized depth *z/h* in Figures 3(a-e) and 4(a-e). It is observed that in the flow region, the stream-wise mean velocity $\overline{u}\big/u_*$ profiles show log-law (Figures 3a,b) throughout the depth, whereas in the blocking (Figure 3c) and wave regions (Figure 3d,e), the mean stream-wise velocity shows log-law away from the boundary (z/h ≥ 0.17), with almost a zero value at the near-bed region. The vertical profile of the stream-wise velocity significantly changes from upstream to downstream with approximately zero velocity at level ~ 0.17 in the wave region. It is interesting to note that at the near-bottom of wave-dominated region, there is an oscillatory nature in the velocity (Figure 3e). This deviation may be due to the strong influence of bed forms on the flow field. Figures 3(d, e) represent the phase-averaged velocity $\dfrac{\langle u \rangle}{u_*}$ due to combined wave-current flows at the downstream. The friction velocity, $u_*$, and equivalent bed roughness, $z_0 = \dfrac{k_s}{30}$, for the wave-blocking condition at five locations ($x/h$ = 28, 42, 44, 47, 50) at a discharge $Q = 0.0197\ m^3/\mathrm{sec}$ over the deformed sediment bed along with only flow over the rigid flat bed are shown in Table-2. It is observed that the equivalent bed roughness $z_0 (= \dfrac{k_s}{30})$ increases from upstream to downstream, that is, from the flow region to the wave region.

**Figure 3     Figure 4**

**Table-2**

The bottom-normal velocity exhibits some interesting behaviours. In the wave dominated region x/h = 47 and 50 (Figures 4d, e), the normalized phase-averaged velocity $\dfrac{\langle w \rangle}{u_*}$ is zero except near the bed, where the velocity shows oscillatory behaviour with negative and positive values; then, from the blocking region to flow region, the bottom-normal velocity shows the reverse phenomenon. As approaching the upstream, the bottom-normal velocity $\dfrac{\overline{w}}{u_*}$ is initially positive near the bed and negative away from the bed; then, it becomes positive again, indicating some internal oscillatory effect of the waves along the depth propagating upstream (Figures 4a-c). It is interesting to note that near the bed, the velocity $\dfrac{\langle w \rangle}{u_*}$ is negative at the wave-dominated region, and as it approaches upstream through the transitional region, the velocity ($\dfrac{\overline{w}}{u_*}$) becomes almost zero at the blocking region, and further upstream at the flow region the velocity is positive.

## Turbulence intensities and Reynolds shear stress

The stream-wise ($\sigma_u$) and vertical ($\sigma_w$) mean turbulence intensities are defined as follows:

$$\sigma_u = \sqrt{\overline{u'^2}} = \sqrt{\dfrac{1}{n}\sum_{i=1}^{n}(u_i - \overline{u})^2} \tag{19}$$

$$\sigma_w = \sqrt{\overline{w'^2}} = \sqrt{\frac{1}{n}\sum_{i=1}^{n}(w_i - \overline{w})^2} \tag{20}$$

The time-averaged Reynolds shear stress component is defined as follows:

$$\tau_{xz} = -\rho\overline{u'w'} = -\frac{\rho}{n}\sum_{i=1}^{n}(u_i - \overline{u})(w_i - \overline{w}) \tag{21}$$

The normalized turbulence intensities $(I_u, I_w)$ and the shear stress $(\tau_{uw})$ are given by the following:

$$I_u = \sigma_u / u_*, I_w = \sigma_w / u_*, \tau_{uw} = -\overline{u'w'} / u_*^2 \tag{22}$$

where $u_*$ is shear velocity determined from the log-low. In the case of wave-induced motion, the normalized phase-averaged stream-wise and bottom-normal intensities are given by the following:

$$\langle I_u \rangle = \frac{1}{N}\sum_{i=0}^{N}\sqrt{(u(t+i\Delta t) - \langle u(t+i\Delta t)\rangle)^2} / u_* \tag{23}$$

$$\langle I_w \rangle = \frac{1}{N}\sum_{i=0}^{N}\sqrt{(w(t+i\Delta t) - \langle w(t+i\Delta t)\rangle)^2} / u_* \tag{24}$$

where $N$ = total number of observations and the normalized phase-averaged Reynolds shear stress is defined as follows:

$$\langle \tau_{uw} \rangle = \frac{1}{N}\sum_{i=0}^{N}(u(t+i\Delta t) - \langle u(t+i\Delta t)\rangle)(w(t+i\Delta t) - \langle w(t+i\Delta t)\rangle) / u_*^2 \tag{25}$$

The phase-averaging of the velocity signals was used to calculate the turbulence parameters at the wave-dominated region (Umeyama, 2009). At first, the time series of the raw velocity signals were checked for waves of a sinusoidal nature. It is observed that at the wave dominated region, the sinusoidal nature of the wave was prominent, and it was progressively reduced towards the bed, and the amplitudes of the waves also reduced to zero at the wave-blocking region.

The stream-wise and bottom-normal intensities are respectively plotted against vertical depth ($z/h$) in Figures 5 and 6 along the flume at the flow region ($x/h$ = 28 and 42), the blocking region ($x/h$ = 44) and the wave region ($x/h$ = 47 and 50). It is observed that the stream-wise intensity $\langle I_u \rangle \approx 1$ throughout the depth at $x/h$ = 50 in the wave region and it decreases until it reaches the blocking region at $x/d$ = 44; then, it increases again at the flow region. A similar trend is also observed for the bottom-normal intensity $\langle I_w \rangle$ as the stream-wise intensity $\langle I_u \rangle$ at all locations (Figure 6), with a smaller magnitude of approximately 0.5. The normalized stream-wise intensity $\langle I_w \rangle$ is dampened near the bottom as it approaches the blocking region.

**Figure 5, Figure 6, Figure 7**

Figure 7 shows the plots of the Reynolds shear stress (RSS) against the depth (*z/h*) at all locations along the flow. RSS defines the momentum flux of the stream- wise velocity in the vertical direction caused by the velocity fluctuations. At the near-bed of the wave-dominated region, the normalized shear stress $\langle \tau_{uw} \rangle$ is nearly zero, and away from the bed, it is positive. It reaches a maximum value at *z/h* = 0.25, corresponding to the maximum momentum transfer, then reduces to zero at the surface. At near-bed of wave-blocking region, the normalized shear stress $\tau_{uw}$ is negative, then positive and reaches a maximum value approximately 2 at the level *z/h* = 0.17, and then again it decreases towards the surface, indicating the changes of momentum flux. At the flow region (a-b), $\tau_{uw}$ increases from zero at the near-bed to a maximum value at the level *z/h* = 0.25, and then reduces towards the surface, which corresponds to the changes of momentum.

## Power spectral density (PSD)

The turbulence spectra provide information on the temporal or spatial scales of velocity fluctuations, from which we can determine the frequency or wave number sub-ranges over which the turbulence is isotropic. To understand the internal structure of the turbulence, the power spectral density (PSD) of the velocity data collected from several locations (*x/h* = 28, 42, 44, 47, 50) was computed using the PSD algorithm with a Hamming Window available in the Horizon ADV software package (Venditti and Bennett 2000; Barman et al. 2016). The filtered velocity data were divided into ensembles of 1024 data points and, hence, 7 ensembles of each of the time series were obtained. All of them were detrended about their means separately. The PSDs are plotted in Figure 8 against the spectral frequency (*f*) in the log-log scale for three selected vertical levels (*z/h* = 0.5, 0.36, 0.04). Here, the rows represent different levels (*z/h*) and the columns represent the five locations (*z/h*). The power spectra of the velocity signal at a high spectral frequency range suggest a good fit to the Kolmogorov - 5/3 power law with a moderately large inertial sub-range at the levels $z/h = 0.5$ and 0.36. At the wave-dominated region near the surface, the PSD peaks at approximately 1 Hz, which decreases towards wave-blocking region and it diminishes at the flow region. The energy spectra at all four downstream locations are very similar. For the near-bed region at $z/h = 0.04$, the power spectra become flat, and the slopes are very small. It is obvious that there is a difference between the near-bed and away from the bed levels, especially at the lowest level $z/h = 0.04$, where the flow is not isotropic due to the generation of bed forms.

**Figure 8**

## Turbulent kinetic energy ( $f_k$ )

The turbulent kinetic energy per unit volume ( $f_k$ ) is estimated using the following equation:

$$f_k = 0.5(\overline{u'^2} + \overline{v'^2} + \overline{w'^2})$$

(26)

for the flow and blocking regions, and the phase-averaged $< f_k >$ case is as follows:

$$< f_k > \ = 0.5 \ (\overline{<u'^2>} + \overline{<v'^2>} + \overline{<w'^2>})$$

(27)

for the wave-dominated region (Raupach, 1981; Venditti and Bauer 2005; Umeyama, 2009). The values of $f_k$ and $< f_k >$ are shown in Table-3 for all five locations ($x/h = 28, 42, 44, 47, 50$) along the flow at three levels ($z/h = 0.5, 0.36, 0.04$). The kinetic energy $f_k$ represents the energy extracted from the mean flow by the motion of turbulent eddies (Kline *et al.*, 1967). It is observed that the overall trend of $f_k$ decreases from the wave region to the flow region (Table-3).

**Table-3**

## Bed form features due to wave-blocking

The signatures of the evolution of the bed forms from upstream to downstream along the flume are shown in Figures 9 (a, b) for the asymmetric ripples beneath the flow region with transitional pattern of current-ripples towards blocking, in Figure 9 (c) for the flat sand bars below the wave-blocking region, and in Figure 9 (d) for nearly symmetric ripples below the wave-dominated region. This variation along the whole sand bed form depends on the different flow regimes generated due to the wave-blocking condition (see the discussion section). Figures 10(a-c) show the side views of the following schematic diagrams, as follows: (a) flat sand bed at the flume prior to the flow, (b) small asymmetric sand waves, which are called the ripple bed, along the flume due to unidirectional flow, and (c) three distinct bed form patterns along the flume bed due to wave-blocking at the water surface.

**Figure 9, Figure 10**

## Statistical analysis of sand bed evolutions

The actual bed form elevations against the longitudinal distance along the centreline of the flume are plotted in Figure 11a for $Q = 0.0183\, m^3/\sec$ and in Figure 11b for $Q = 0.0197$ $m^3/\sec$. This clearly indicates that the elevations of asymmetric ripples on the left at the flow-dominated region are higher than that of the rippled beds developed on the wave-blocking and wave-dominated regions. Particularly, the bed elevations over the flat sand bars and symmetric ripples are almost identical, except for one spike (outlier) at the wave region (Figure 11a). Therefore, the plots suggest that there is a significant change in the bed forms due to wave-blocking. Three distinct regions are indicated in Figures 11(a, b).

It is observed that the recorded bed elevations along the modified bed form for each pair of discharges ($Q$) and the corresponding wave frequencies ($\omega_c$) change randomly with space. To understand the variations of the bed elevation over the entire bed forms from upstream to downstream, the bed elevation data are analysed for four different discharges with the corresponding frequencies. The statistical parameters, i.e., the mean and the standard deviation or r.m.s. values, of the bed elevations are computed for three distinct regions along the entire bed forms (Table-4). The table shows that the statistical parameters change significantly in the respective regions.

**Figure 11**

The recorded elevation data from upstream to downstream show three distinctly different regions with positive and negative bed elevations. The positive elevation increments ($> 0$) correspond to deposition ($D$), the negative increments ($< 0$) correspond to scour ($S$) and the zero values of increment ($= 0$) correspond to inactivity at that location. Each of these three processes, i.e., deposition, scour and inactivity, has a characteristic elevation scale of operation. The probabilistic aspect of the bed form evolution is performed along the bed for the three regions, when the stable bed forms occurred during the total 240 minutes of time.

Following Ganti et al. (2011) and Sarkar et al. (2016), the exceedance probability of the series of deposition $D$ for the deposition magnitude $D_j$ is defined as follows:

$$P(D > D_j) = j/(m+1) \tag{28}$$

where $j$ corresponds to the rank of the deposition in the sorted series of $D$ in descending order, e.g., $j =1$ for the largest deposition magnitude in the series of $D$ and $m$ is the length of $D$ series, i.e., equal to the maximum value of $j$. Similarly, the exceedance probability $P(S > S_j)$ of the series of scour event $S$ for the scour magnitude $S_j$ is calculated.

**Table-4**

Here, log-log plots of exceedance probabilities $P(D > D_j)$ and $P(S > S_j)$ for all three distinct regions are shown in Figures 12 (a, b) for $Q = 0.0183\, m^3/\text{sec}$ and in Figures 13 (a, b) for $Q = 0.0197\, m^3/\text{sec}$, including the entire region from upstream to downstream. In fact, the exceedance probability is the probability that a certain depositional height or scour depth recorded in a location of stable state will be exceeded during the whole experimental data along a certain region. From the figure, it is clearly noted that the overall minimum exceedance probabilities of deposition $P(D > D_j)$ and scour $P(S > S_j)$ events are observed for the wave-dominated region, whereas the maximum exceedance probabilities of both deposition and scour events are observed for the upstream flow region in Figures 12(a, b). It is clearly visible that the exceedance probability of deposition $P(D > D_j)$ at a certain elevation $D$ in the wave-blocking region is greater than that of wave-dominated, and less than that of upstream flow as well as the entire elevation region along the flume. The bed elevations (positive or negative) at the upstream flow and the downstream counter-wave regions show significant changes in the bed form development. The exceedance probabilities of deposition and scour events at both wave-blocking and wave-dominated regions are identical, whereas the magnitudes of elevations (positive or negative) at wave-blocking are greater than that of the counter-wave region.

Moreover, the probability $P(D > D_j)$ is greater at the wave-blocking region for a magnitude of $D = 1.0$ cm than that for $P(S > S_j)$ for the scour event. It is also observed from the figures that the maximum magnitudes of the deposition event $D$ in the upstream flow and total flow regions are approximately 3 times greater than that of the scouring event S. A similar nature is also observed for $Q = 0.0197 \ m^3/\text{sec}$ from Figures 13(a, b). It is observed that the bed elevations (positive or negative) for $Q = 0.0197 \ m^3/\text{sec}$ are greater than that for $Q = 0.0183 \ m^3/\text{sec}$.

**Figure 12,  Figure 13**

## Discussions

Flume experiments were conducted in a laboratory on the propagation of surface waves against the flow over a sediment bed of median particle diameter $d_{50} = 0.25$ mm. Under certain critical frequencies of waves in a particular flow discharge, a wave-blocking phenomenon in a stream-wise location was observed, even over the sediment bed. The novelty of this work was to recognize the wave-blocking due to counter propagating waves and its impact on the growth of bed forms. The non-uniformity of flow along the flume with three distinct flow patterns was noticed. Owing to this non-uniform flow condition, the sediment bed was segmented into three different regions of bed geometry.

The experimental studies were performed in the three following steps:

(i)     The fully developed flow over the rigid bed surface was verified. The dimensionless stream-wise mean velocity over the flat rigid surface followed the log-law at all the measuring locations. Apart from this, the normalized turbulence intensities in the stream-wise and vertical directions above the rigid flat surface agreed well with those of Nezu and Rodi (1986), Clifford, (1998), Ojha and Mazumder, (2010), and Chatterjee et al. (2018), with slight changes of empirical coefficients. Yang et al. (2004) and Absi (2011) reported that a two-dimensional flow in the central portion of the flume was achieved with a width/depth ratio $\approx 2$. In the present study, the width/depth ratio was $\approx 2.7$ with the occurrence of the maximum velocity at the level $H = 18$ cm as a dip-phenomenon, so it could be ascertained that the two-dimensional flow at the central portion of the flume and, hence, the effect of secondary currents due to sidewalls was negligible. Moreover, the zero vertical velocity throughout the depth also confirmed that a fully developed flow was achieved.

(ii)    After a certain time of uniform flow over the flat sand bed, asymmetrical ripples of sand with steeper slopes towards the downstream were formed (Allen, 1968; Venditti et al. 2005; and others). The ripple dimensions, such as amplitude and wavelength, over the bed forms were dependent on the flow discharge and the bed characteristics. The ripples played an important role in generating the flow separation, turbulence, flow resistance and controlling the sediment transport over two-dimensional bed forms.

(iii) The wave-blocking condition over the sand bed of asymmetrical ripples was created using a predetermined critical frequency of a wave in a particular discharge to examine the changes in the existing bed forms. After an equilibrium state of modified bed forms occurred, the velocity data were recorded to enumerate the modulated turbulence parameters due to wave-blocking over the bed forms.

The appearance of wave-blocking due to counter-propagating waves was verified using the linear dispersion relation for low amplitude waves. The wave action equation proposed by Bretherton and Garrett (1968) was used to explain the wave-blocking phenomenon. The wave-blocking appeared at a point when the absolute group velocity at a stationary frame vanished, which followed the ratio of the depth averaged velocity over the sediment bed and the group velocity ~ 1, indicating the formation of wave-blocking over the sediment bed. However, several relevant features, i.e., non-linearity and frequency downshifting (Ma et al. 2010; and Shugan et al. 2015) to the counter propagating waves, are not considered in the present study.

Due to the non-uniformity in the flow, the variations of the mean velocity components along the longitudinal and bottom-normal directions are discussed. In fact, below the wave-dominated region at near-bottom, the oscillatory nature of velocity was observed, which could be due to the strong influence of bed forms on the flow structure. To accurately describe the two-dimensional flow over bed forms requires the application of models more robust than the log-law, such as those introduced by McLean and Smith (1986) and Nelson and Smith (1989). It is interesting to note that due to the wave-blocking condition at the near-bed region, the bottom-normal velocity showed an oscillatory behaviour from down- to upstream with negative to positive values. The oscillating behaviour of $\overline{w}/u_*$ around zero at the near-bed region was mostly responsible for generating the symmetric ripples at the wave-dominated region. This phenomenon was also confirmed from an oscillatory nature of stream-wise velocity $\langle u \rangle/u_*$ at the near-bed region (Figure 3e).

The equivalent bed roughness $z_0 (= \frac{k_s}{30})$ increased from upstream to downstream, that is, from the flow region to the wave region. The strong influence of bed roughness towards the wave region was demonstrated by the presence of local non-homogeneity of the bed roughness due to the wave against the current. In this case, bed roughness was produced mainly from the bed forms developed by the counter wave over the sand bed, and roughness due to skin friction was considered negligibly small. It is also observed that the friction velocity $u_*$ showed a non-uniform behaviour along upstream to downstream, which might be due to wave-blocking condition interacting with the sediment bed.

It is noted that the overall trends of both the intensity components (longitudinal and vertical) are less at the blocking region compared to the flow and wave regions, which may be due to the control of the eddy movement. A zone of higher turbulence intensity is approved to a high level of noise from the acoustic reflection of the bed (Nikora and Goring, 1998) and is characterized by low mean velocity. The careful examination of Figure 7 shows that the overall trend of the Reynolds shear stress (RSS) is negative near the bottom at the blocking and wave-dominated regions, which results in the outward flux of momentum and in positive values at the near surface level resulting in the inward flux of momentum. This is consistent with the results of Maity and Mazumder (2014). In the flow region, the RSS is positive, indicating an inward flux of momentum.

According to Corvaro et al.'s (2014) observations on spectral density, approaching the sediment bed from water surface, the PSD shifted towards higher frequencies and energy attenuation became larger. They also observed that PSD was closer to -5/3 Kolmogorov's law near the free surface and became flatter near the bed. Similarly, in the present case, the PSDs near the water surface and mid-depth approximately fitted to the -5/3 law and became flatter near the bed; therefore, the slopes are very small near the bed. The flatness was greater in wave region compared to the other regions. The spectra of velocity would be identical at all levels if the flow was completely isotropic. The velocity power spectra showed that the surface energy was maximal at the wave-dominated region, and it reduced towards the wave-blocking region, and finally, no frequency peak occurred at the flow region. The surface energy diminished gradually from the wave region to the flow region. Moreover, the turbulent kinetic energy decays from the wave to the flow region. This trend showed in all the vertical profiles.

The effect of wave-blocking on the asymmetric bed forms showed a transitional pattern and suggested that the bed forms could be segmented into three regions, as follows:

(i)      In the flow region, the asymmetric ripples for uniform flow prevail. The ripples are sharper on the wave side and slope more gently on the flow side.

(ii)     Beneath the wave-blocking region, the sand ripples are more disorganized and fewer ripples occur, which are of lower height, similar to flat sand bars.

(iii)    In the wave-dominated region, the ripples are nearly symmetrical due to the interaction between the counter flow and wave forces.

In the flow regime, asymmetric sand waves with steeper slopes in the downward face, called ripples, are sustained. The ripple length depends on the sediment size but is essentially independent of the water depth (Figure 9a). Figure 9b shows the transitional pattern of current-ripples towards the wave-blocking region. The transition from the ripple bed to the flat sand bed depends on the interactions of velocity and opposing wave motion. In the wave-blocking region, the modification from an asymmetric rippled bed to flat sand bed form are possibly due to the dissipation of wave energy, and hence, there is drastic drop in turbulent fluctuations and resistance throughout the depth of flow (Figure 9c). Thus, it can be stated that counter-propagating waves modulate the large-scale velocity fluctuations within the blocking region from the top to the near-bed surface. The formation of nearly symmetric ripples beneath the wave-dominated region may be due to the generation of periodic waves, where the volume of sediment moves back and forth during the wave period (Figure 9d), and hence, symmetric bed forms are observed.

According to Dean and Dalrymple (2000) the wave characteristics are found to be that of deep water and intermediate deep water for the present case (Table 1c), though effectively there is no orbital motion in the bed to form the ripples. However, the formation of different patterns in the bed forms depends mostly on the interaction between the wave and counter-current, particularly flat sand bars in the wave-blocking and symmetric ripples in the wave-dominated regions. The symmetric patterns in ripples may be formed from the effect of the oscillatory nature of velocities (Figures 3 and 4) and shear stress at the near-bed region (Figure 7), even in the deep water flow. Although the formation of bed forms is not directly affected by orbital motion, there is an interaction between the waves and the counter current with the sediment bed, which is responsible for generating the oscillatory nature of flow parameters at the near-bed region; hence, there is a formation of symmetric bed forms beneath the wave-dominated region. According to Ribberink (1995), to increase the opposing current, higher wave velocity values are necessary to transform a ripple bed to a plane bed, whereas for opposing current and lower values of wave velocities, the reverse result takes place. The present findings over the sediment bed due to the wave-blocking phenomenon, to some extent, corroborate to his results.

Wave-induced symmetric ripples indicate an environment of weak currents, where the flow is dominated by wave oscillations. In fact, wave-induced ripples are traditionally described as symmetrical. It is interesting to note that the interaction of counter propagating waves and the bottom sand waves results the bed form patterns. Therefore, Figure 9 shows the novelty of the present work in the development of bed forms due to the wave-blocking phenomenon. It is apparent that the entire sediment bed along the flume may be designated into three distinct patterns.

The statistical analysis shows that the standard deviations of the bed elevations in the wave and blocking regions are significantly reduced for two higher discharges compared to that of the flow region, indicating the uniformity of intensities of fluctuations in elevations; whereas for smallest discharge, the standard deviations are almost identical, indicating the uniformity of turbulence along the sand bed. The exploratory data analysis suggests that there is a change in the bed form elevations along the flume with asymmetric ripples with steeper slopes at the downstream face, flat sand bars in the wave-blocking region, and almost symmetric ripples below the wave region, which may be segmented into three regions, such as, the flow region at the entrance, the wave-blocking region, and the wave region at the downstream, respectively. This effect is relevant to understand the development of bed form patterns when there is a wave-blocking on the water surface.

The exceedance probabilities of deposition $P(D > D_j)$ and scour $P(S > S_j)$ events are minimal for the wave region, but the probabilities of both events are maximal for the upstream flow region. Additionally, the exceedance probability of $P(D > D_j)$ at a certain D at the wave-blocking region is greater than that of wave-dominated region, and less than that of the flow region as well as the total elevations along the flume. From our findings, it is inferred that the deformed bed tends to flatten gradually from the flow dominated region to the wave-blocking region (Ribberink 1995). The changes in the bed elevation (deposition or erosion) process are random in nature, the log-log plots of the exceedance probability of deposition and scour help to define the thickness of the preserved stratigraphic sequences (Kolmogorov, 1951; Paola et al., 2009).

## Conclusions

The novelty of the experiment is to identify the wave-blocking for counter-propagating waves with the flow over a sediment bed and to investigate the effect of wave-blocking on the sediment bed. Wave-blocking is formed for a particular pair of discharge and the frequency of the counter-propagating wave, where the amplitudes of the waves diminish, and the current reaches the group velocity. Therefore, it modulates the large-scale velocity fluctuations within the blocking region from the top to the near-bed surface. Regarding the sand bed, when the frequency of waves is superimposed on the current, the interaction of the wave and current produces a dramatic diffusion of the sediment layer at the bottom boundary. The effect of wave-blocking over the asymmetric bed forms shows a transitional pattern, and the bed forms may be segmented into three distinct regions.

According to Ribberink (1995), higher wave velocities are necessary to transform a ripple to a plane bed for higher opposing current, whereas lower wave velocity values are necessary to transform the ripple into the plane bed for an increase in the following current. The velocity power spectra show that the surface energy is maximal at the wave-dominated region, and it reduces towards wave-blocking, and finally, the frequency peak vanishes at the flow region. In the near-bed region, power spectra become flat, so the slopes are very small. The surface energy diminishes gradually from the wave region to the flow region. Moreover, the turbulent kinetic energy decays from the wave to the flow region.

The paper has contributed a framework for understanding the turbulence properties of the wave-current flow over a loose sediment bed when waves propagate against the flow with the wave-blocking condition. The knowledge of such turbulence parameters and the associated interactions with the sand bed may be applicable to coastal and river engineering problems, relating to the dynamics of flow and bed forms at river-mouths and tidal inlets. It can also help in the understanding of preserved stratigraphic sequences from the depositional and scouring events due to wave-blocking over sediment beds for the interpretation of palaeoenvironmental conditions (Kolmogorov, 1951; Sengupta, 2007; Paola et al., 2009).

**Acknowledgements -** It is indeed a pleasure to thank Prof. Silke Weinfurtner for the suggestions regarding the experimental setup. We must acknowledge Prof. J. T. Kirby for providing us his important papers on the wave-blocking phenomena. The authors are also grateful to Prof. Koustuv Debnath, Dr. Satya Praksh Ojha and Dr. Haradhan Maity for their helpful discussions. Our thanks are extended to two anonymous reviewers and the Associate

Editor for their concrete comments and suggestions for the paper and to Prolay Das for his help during the experiments. Finally authors would like to acknowledge The Wiley through http://wileyeditingservices.com for editing the language in the manuscript. Authors do not have any conflict of interest to declare regarding the paper.

**Table -1.** Summary of the parameters for all experimental conditions.

<u>(a) Step-1: Flow only over a rigid flat bed</u>

| $Q$ | $Q_{cal}$ | $H$ | $u_m$ | $u_*$ | $u_{avg}$ | $k_s$ | Re | $Fr$ | Bed conditions |
|------|------|------|------|------|------|------|------|------|------|
| (m³/s) | (m³/s) | (cm) | (cm/s) | (cm/s) | (cm/s) | (cm) | | | |
| 0.0197 | 0.0309 | 25 | 24.73 | 0.91 | 18.37 | 0.025 | $0.62 \times 10^5$ | 0.158 | rigid flat bed |
| 0.0322 | 0.0511 | 25 | 40.84 | 1.59 | 32.38 | 0.036 | $1.02 \times 10^5$ | 0.261 | rigid flat bed |

Reynolds number $\mathrm{Re} = u_m H \big/ \nu$ and Froude number $Fr = \dfrac{u_m}{\sqrt{gH}}$, $Q$ is the recorded flow discharge, $Q_{cal}$ is the calculated flow discharge, $u_*$ is the shear velocity, $u_m$ is the maximum velocity, $u_{avg}$ is the depth averaged velocity, and $k_s$ is the equivalent bed roughness.

<u>(b) Step -2: Flow only over a sand bed of median particle size $d_{50} = 0.25$ mm</u>

| $Q (m^3/s)$ | $h$ (cm) | $h'$ (cm) | $u_m$ (cm/s) | Bed conditions |
|------|------|------|------|------|
| 0.0169 | 15 | 10 | 31 | Asymmetric |
| 0.0183 | 15 | 10 | 34 | rippled sand bed |
| 0.0197 | 15 | 10 | 37 | |
| 0.0208 | 15 | 10 | 39 | |

<u>(c) Step-3: Wave-blocking conditions over a sand bed in a wave-dominated region</u>

| $Q (m^3/s)$ | $\omega_c$ (Hz) | h (cm) | $\lambda$ (cm) | $h/\lambda$ | $kh (=2\pi h/\lambda)$ | Wave nature | Bed conditions |
|------|------|------|------|------|------|------|------|
| 0.0169 | 1.23 | 15 | 23 | 0.65 | 4.08 (> $\pi$) | deep water | Three different |
| 0.0183 | 1.15 | 15 | 25 | 0.60 | 3.76 (> $\pi$) | deep water | bed forms |
| 0.0197 | 1.10 | 15 | 28 | 0.53 | 3.33 (> $\pi$) | deep water | observed along |
| 0.0208 | 1.01 | 15 | 33 | 0.45 | 2.82 (< $\pi$) | intermediate | the flume |

**Table-2.** Values of the flow parameters in different regions for the wave blocking condition at a discharge Q = 0.0197 $m^3/\sec$.

| Only flow over a rigid flat surface | | Wave-blocking condition over a sediment bed | | | | | | | | | |
|---|---|---|---|---|---|---|---|---|---|---|---|
| | | \multicolumn{12}{}{Discharge $Q = 0.0197$ $m^3/\sec$} |
| | | Flow region | | | | Wave-blocking region | | Wave-dominated region | | | |
| | | $x/h = 28$ | | $x/h = 42$ | | $x/h = 44$ | | $x/h = 47$ | | $x/h = 50$ | |
| $u_*$ | $z_0$ | $u_*$ | $z_0$ | $u_*$ | $z_0$ | $u_*$ | $z_0$ | $u_*$ | $z_0$ | $u_*$ | $z_0$ |
| (cm/s) | (cm) | (cm/s) | (cm) | (cm/s) | (cm) | (cm/s) | (cm) | (cm/s) | (cm) | (cm/s) | (cm) |
| 0.91 | 0.00084 | 3.702 | 0.106 | 3.169 | 0.121 | 4.08 | 0.251 | 4.38 | 0.36 | 4.036 | 0.37 |

**Table -3.** Turbulent kinetic energy ($f_k$) at different horizontal locations and vertical levels at flow discharge Q = 0.0197 $m^3/\sec$ .

| Vertical height | Flow region | | Wave-blocking | Wave-dominated region | |
|---|---|---|---|---|---|
| z/h | x/h = 28 | x/h = 42 | x/h = 44 | x/h = 47 | x/h = 50 |
| | $f_k$ (erg) | $f_k$ (erg) | $f_k$ (erg) | $<f_k>$ (erg) | $<f_k>$ (erg) |
| 0.50 | 7.47 | 16.72 | 17.50 | 26.15 | 37.84 |
| 0.36 | 16.44 | 20.86 | 24.36 | 28.09 | 25.60 |
| 0.04 | 55.97 | 63.41 | 99.85 | 52.56 | 158.78 |

**Table-4.** Statistical parameters: the mean and standard deviation or r.m.s. values of the bed elevations for three different regions and four discharges, including the only flow condition.

| | Deformation of bed forms in wave-blocking condition | | | | | | Deformation of bed forms in the only flow condition | |
|---|---|---|---|---|---|---|---|---|
| Flow discharge ($m^3/s$) | Wave-blocking region | | Flow region | | Wave region | | Mean | Standard deviation |
| | Mean (cm) | Standard deviation (cm) | Mean (cm) | Standard deviation (cm) | Mean (cm) | Standard deviation (cm) | (cm) | (cm) |
| 0.0208 | 6.3269 | 0.5205 | 7.1064 | 1.3480 | 6.7621 | 0.5744 | 6.8091 | 1.0962 |
| 0.0197 | 6.8413 | 0.7690 | 7.3256 | 1.7491 | 6.8365 | 0.5425 | 8.3010 | 0.8419 |
| 0.0183 | 6.4794 | 0.4249 | 6.8560 | 1.4154 | 7.0100 | 1.2608 | 8.0356 | 0.7111 |
| 0.0169 | 7.3173 | 0.5381 | 7.4950 | 0.5536 | 7.2564 | 0.6033 | 8.5549 | 1.1738 |